\documentclass[10pt,superscriptaddress,twocolumn,amsmath,amssymb,aps,pra]{revtex4-1}
\usepackage{mathrsfs}
\usepackage{graphicx}
\usepackage{dcolumn}
\usepackage{bm}
\usepackage{hyperref}

\newcommand{\be}{\begin{equation}}
\newcommand{\ee}{\end{equation}}
\newcommand{\bea}{\begin{eqnarray}}
\newcommand{\eea}{\end{eqnarray}}

\begin{document}
\title{Observability of Higgs Mode in a system without Lorentz invariance}
\author{Xinloong Han}
\affiliation{ Institute of Physics, Chinese Academy of Sciences,
Beijing 100190, China}

\author{Boyang Liu } \email{boyangleo@gmail.com} \affiliation{Department of Physics and Center of Theoretical and Computational Physics,
The University of Hong Kong, Hong Kong, China}

\author{Jiangping Hu  } \affiliation{
Institute of Physics, Chinese Academy of Sciences, Beijing 100190,
China}\affiliation{Department of Physics, Purdue University, West
Lafayette, Indiana 47907, USA}

\date{\today}
\begin{abstract}
We study the observability of the Higgs mode in BEC-BCS crossover.  The observability of Higgs mode is investigated by calculating the spectral weight functions of the amplitude fluctuation below the critical transition temperature. At zero temperature, we find that there are two sharp peaks on the spectral function of the amplitude fluctuation attributed to Goldstone and Higgs modes respectively. As the system goes from BCS to BEC side,  there is strong enhancement of spectral weight transfer from the Higgs to Goldstone mode. However, even at the unitary regime where the Lorentz invariance is lost,  the sharp feature of Higgs mode still exists. We specifically calculate the finite temperature spectral function of amplitude fluctuation at  the unitary regime and  show that the Higgs mode is observable at the temperature that present experiments can reach.

\end{abstract}
 \maketitle

\section{Introduction}
A general description of many-body theory shows that after a spontaneous symmetry breaking two kinds of collective modes emerge \cite{Pekker2014}. One is known as the Goldstone mode originated from the phase fluctuation of order parameters. The other one is referred as ``Higgs mode", which is from the amplitude fluctuation of the order parameter. With the triumph of the observation of Higgs particle in high energy physics~\cite{Cern1, Cern2}, the interests of Higgs collective modes in many-body systems are growing. The earliest observation is in the Raman scattering experiment of a superconducting charge density wave compound $\text{NbSe}_2$ in 1980s'~\cite{Sooryakumar, Littlewood1, Littlewood2}. Recently, with the development of experimental techniques more evidences of the Higgs mode in many-body systems have appeared. For example, Higgs mode has been observed in antiferromagnet $\text{TlCuCl}_3$ by the neutron scattering~\cite{Ruegg}, and the terahertz pump probe spectroscopy also revealed signal of Higgs mode in superconducting NbN sample in a nonadiabatic excitation regime~\cite{Matsunaga1, Matsunaga2}. In cold atom system, Higgs mode has been observed in bosonic atoms in optical lattices when fine tuning the parameters to the Lorentz invariance point ~\cite{Endres}. Various theoretical researches have also been conducted \cite{Varma2002,Barlas2013,Huber2008,Pollet2012,Podolsky2011,Podolsky2012,Gazit2013,Rancom2014,Liu,Cea}.

Usually, the existence of Higgs mode requires the Lorentz invariance. In ref. \cite{Liu} one of us has manifestly discussed how the Higgs mode evolved as the system is tuned away from the Lorentz invariance point. Basically, as the Lorentz invariance is gradually lost, the spectral weight transfer from the Higgs mode to the Goldstone mode is enhanced; meanwhile, the Higgs mode becomes more strongly coupled to the quasi-particle excitations and other collective modes and finally gets overdamped. Hence, close to the unitary regime, the Higgs mode is non-observable. However, in ref. \cite{Liu} the whole derivation is based on the time-dependent Ginzburg-Landau theory. The validity is constrained in the vicinity of the transition temperature, where the damping effect is strong compared with the case of lower temperature. It is interesting to investigate the observability of Higgs mode using a mean-field theory in the case deep below the critical temperature.

In this work we study the Fermion superfluid of dilute Fermi gas with tunable interaction. In our theoretical framework the superfluid order parameter is represented as mean-field plus the phase and amplitude fluctuations. The spectral weight functions of the phase and amplitude fluctuations can be calculated using the functional integral method. First, we discuss the spectral weight transfer of the Higgs mode and Goldstone mode at zero temperature and show that the signal of Higgs mode is still clear in the unitary regime even though the Lorentz invariance is lost here. Second, we investigate the observability of the Higgs mode as the function of temperature in the unitary regime since the superfluid transition temperature in this region is reachable in the present experiments.

The paper is organized as follows. In Sec.\ref{sec:mf}, we describe the theory that is employed to investigate the problem. In Sec.\ref{sec:T0}, we calculate the spectral functions of the amplitude and phase fluctuations and discuss the spectral weight transfer of the Higgs and Goldstone modes. In Sec.\ref{sec:T}, we calculate the spectral function of the amplitude fluctuation at finite temperature and demonstrate the observability of Higgs mode in the unitary regime. Finally, we give a brief conclusion in Sec.\ref{sec:con}.

\section{Path integral description of the BEC-BCS crossover\label{sec:mf}}
The partition function that describes a system of two species fermions can be written as $\mathcal
Z=\int
D[\bar\psi_\sigma,\psi_\sigma]e^{-S[\bar\psi_\sigma,\psi_\sigma]}$,
with the action \be S[\bar\psi_\sigma,\psi_\sigma]=\int d\tau d^3{\bf x}\Big
\{\bar\psi_\sigma(\partial_\tau-\frac{\nabla^2}{2m}-\mu)\psi_\sigma-g\bar\psi_\uparrow
\bar\psi_\downarrow\psi_\downarrow\psi_\uparrow\Big \}, \ee where
$\psi_\sigma$ are the fermionic fields with spin index $\sigma=\uparrow,\downarrow$. The parameter $g$ is the contact interaction
between fermions of opposite spins. It is related to the s-wave scattering length $a_s$ by $1/g=-m/4\pi a_s+\sum_{\bf k}1/2\epsilon_{\bf k}$ with $\epsilon_{\bf k}={\bf k}^2/2m$. Introducing an auxiliary field $\Delta(\tau, {\bf x})$, which represents the cooper pair field, the four fermion
interaction term can be decoupled in the Cooper channel using the Hubbard-Stratonovich transformation. After integrating out the fermionic fields $\psi_\sigma$
one obtains an effective theory of Cooper pair field
 $\Delta$ as \be \mathcal Z=\int D(\bar\Delta,\Delta)\exp
\Big[-\frac{1}{g}\int d\tau d{\bf x}|\Delta|^2+\ln\det\hat
G^{-1}\Big], \label{eq:Z} \ee where \be \hat{G}^{-1}=
\left(\begin{array}{cc}-\partial_\tau+\frac{\nabla^2}{2m}+\mu &
\Delta\\ \bar \Delta &
-\partial_\tau-\frac{\nabla^2}{2m}-\mu\end{array}\right) \ee
 is the Gor'kov Green function.

To investigate the collective modes in the superfluid state we write the field $\Delta(\tau, {\bf x})$ as \cite{Engelbrecht, Diener}
\bea  \Delta(\tau, {\bf x})=\Delta_0+\delta\phi(\tau, {\bf x}),\eea where $\Delta_0$ is the mean-field and $\delta\phi(\tau, {\bf x})$ represents the fluctuations. we next expand the action up to the second order of fluctuations. In the momentum space the action can be cast as \bea && S[\delta\phi^\ast,\delta\phi]=\cr&&\frac{1}{2}\sum_{i\omega_n}\int \frac{d^3{\bf k}}{(2\pi)^3}\overline{\Phi}(i\omega_n,{\bf k}){\bf M}(i\omega_n,{\bf k})\Phi(i\omega_n,{\bf k}),  \eea where we rearrange the fluctuation in a form of spinor as $\overline{\Phi}(i\omega_n,{\bf k})=[\delta\Phi^\ast(i\omega_n,{\bf k}), \delta\Phi(-i\omega_n,-{\bf k})]$ and $i\omega_n=i2n\pi/\beta$ is the bosonic Matsubara frequency. The inverse propagator ${\bf M}$ is given by \bea && {\bf M}_{11}(i\omega_n,{\bf k})={\bf M}_{22}(-i\omega_n,-{\bf k})=-\frac{m}{4\pi a_s}+\cr&&\int\frac{d^3{\bf q}}{(2\pi)^3}\Bigg\{[1-f-f']\Big(\frac{u^2u'^2}{i\omega_n-E-E'}-\frac{v^2v'^2}{i\omega_n+E+E'}\Big)\cr&&+[f-f']
\Big(\frac{v^2u'^2}{i\omega_n+E-E'}-\frac{u^2v'^2}{i\omega_n-E+E'}\Big)+\frac{1}{2\epsilon_{\bf k}}\Bigg\},\cr&&
{\bf M}_{12}(i\omega_n,{\bf k})={\bf M}_{21}(i\omega_n,{\bf k})=\cr&& \int\frac{d^3{\bf q}}{(2\pi)^3}\Bigg\{[1-f-f']\Big(\frac{uvu'v'}{i\omega_n+E+E'}-\frac{uvu'v'}{i\omega_n-E-E'}\Big)\cr&&+[f-f']
\Big(\frac{uvu'v'}{i\omega_n+E-E'}-\frac{uvu'v'}{i\omega_n-E+E'}\Big)\Bigg\}. \label{eq:M}\eea In above expressions we use the conventional notation $u=u_{\bf q}$, $v=v_{\bf q}$, $E=E_{\bf q}$ and $u'=u_{\bf q+k}$, $v'=v_{\bf q+k}$, $E'=E_{\bf q+k}$, where $E_{\bf q}=\sqrt{\xi_{\bf q}^2+\Delta^2_0}$, $v^2_{\bf q}=1-u_{\bf q}^2=(1-\xi_{\bf q}/E_{\bf q})/2$ and $\xi_{\bf q}=\epsilon_{\bf q}-\mu$. $f=1/[\exp(\beta E)+1]$ and $f'=1/[\exp(\beta E')+1]$ are the Fermi distribution.

\section{The Spectral Weight function of Higgs mode at temperature $T=0$\label{sec:T0}}
To investigate the observability of Higgs mode it is useful to sperate the fluctuation $\delta\phi$ into real and imaginary parts $\delta\phi(\tau,{\bf x})=(\delta_a(\tau,{\bf x})+i\delta_p(\tau,{\bf x}))/\sqrt 2$. Then $\delta_a$ and $\delta_p$ describe the amplitude and phase
fluctuations, respectively. After a rotation the effective action can be written in a form of\bea && S[\delta_a,\delta_p]=\cr&&\frac{1}{2}\sum_{i\omega_n}\int \frac{d^3{\bf k}}{(2\pi)^3}[\delta_a^\ast,\delta_p^\ast]{\bf Q}(i\omega_n,{\bf k})\left[\begin{array}{c}\delta_a\\ \delta_p\end{array}\right],  \eea
where the inverse propagator ${\bf Q}(i\omega_n,{\bf k})$ is expressed in terms of ${\bf M}_{ij}(i\omega_n,{\bf k})$ as\bea {\bf Q}= \left[\begin{array}{cc}{\bf M}^E_{11}+{\bf M}_{12}&i{\bf M}^O_{11}\\
-i{\bf M}^O_{11}&{\bf M}^E_{11}-{\bf M}_{12}\end{array}\right]. \label{eq:Q}\eea
Here we redefine ${\bf M}^E_{11}={\bf M}_{11}+{\bf M}_{22}$ and ${\bf M}^O_{11}={\bf M}_{11}-{\bf M}_{22}$. The index ``E" and ``O" denote even and odd in $i\omega_n$.

In this section we study the spectral weight transfer at zero temperature. At $T=0$ limit the Fermi distribution function $f$ and $f'$ in Eq. (\ref{eq:M}) vanish. To study the dispersion relation and the spectral weight properties of the collective modes one usually take  the analytic continuation $i\omega_n\rightarrow \omega+i0^+$ of the inverse propagator ${\bf Q}(\omega+i0^+,{\bf k})$ and then expand it in terms of small momentum ${\bf k}$ and energy $\omega$ \cite{Engelbrecht} up to the second order. However, in our calculation based on mean-field thoery the expansion in small $\omega$ is not legitimate. While we expand ${\bf Q}(\omega+i0^+,{\bf k})$ in terms of small $\omega$ we actually treat $\omega/{\rm min}\{E_{\bf k}\}$ as a small parameter, where ${\rm min}\{E_{\bf k}\}$ is the minimum value of $E_{\bf k}$. The Higgs mode gap is of order $\Delta_0$. Hence, we are interested in the energy scale of $\omega\sim\Delta_0$. In this case $\omega/{\rm min}\{E_{\bf k}\}$ might not be a small parameter. For example, at the BCS limit ${\rm min}\{E_{\bf k}\}$ is of order $\Delta_0$ around the Fermi surface, then $\omega/{\rm min}\{E_{\bf k}\}\sim 1$. In our work we will directly calculate the spectral weight function from the inverse propagator ${\bf Q}(\omega+i0^+,{\bf k})$ without small $\omega$ and ${\bf k}$ expansion. After taking the analytic continuation $i\omega_n\rightarrow \omega+i0^+$ the matrix elements of the ${\bf Q}(\omega,{\bf k})$ can be explicitly written as\bea &&{\bf Q}_{11}(\omega,{\bf k})=-\frac{m}{4\pi a_s}\cr&&+\int\frac{d^3{\bf k}}{(2\pi)^3}\Big\{\frac{E+E'}{2EE'}\cdot\frac{EE'+\xi\xi'-\Delta_0^2}{(\omega+i0^+)^2-(E+E')^2}+\frac{1}{2\epsilon_{\bf k}}\Big\},\cr&&{\bf Q}_{22}(\omega,{\bf k})=-\frac{m}{4\pi a_s}\cr&&+\int\frac{d^3{\bf k}}{(2\pi)^3}\Big\{\frac{E+E'}{2EE'}\cdot\frac{EE'+\xi\xi'+\Delta_0^2}{(\omega+i0^+)^2-(E+E')^2}+\frac{1}{2\epsilon_{\bf k}}\Big\},\cr&&{\bf Q}_{12}(\omega,{\bf k})={\bf Q}_{21}(-\omega,-{\bf k})=\cr&&i\int\frac{d^3{\bf k}}{(2\pi)^3}\frac{E\xi'+E'\xi}{4EE'}\cdot\frac{\omega}{(\omega+i0^+)^2-(E+E')^2}.\label{eq:q0}\eea

First, we will study the collective modes at BCS limit. At zero temperature the off-diagonal term in Eq. (\ref{eq:Q}) can be calculated as ${\bf Q}_{12}=i{\bf M}_{11}^O=i\int\frac{d^3{\bf k}}{(2\pi)^3}\frac{E\xi'+E'\xi}{4EE'}\cdot\frac{\omega}{\omega^2-(E+E')^2}$. At BCS limit the integrand in ${\bf Q}_{12}$ is approximately an odd function of ${\bf k}$ with respect to the axis $\epsilon_{\bf k}=\mu$. Then when one takes the integration of it the positive and negative part will cancel out each other. That is, the off-diagonal term  ${\bf Q}_{12}$ vanishes. This manifests the particle-hole symmetry at BCS limit. The spectrum of the collective modes can be calculated from the equation ${\rm det}[{\bf Q}(\omega,{\bf k})]=0$. As a result, ${\bf Q}_{11}$ and ${\bf Q}_{22}$ decouple. It's straight forward to see that we have two modes defined by ${\rm Re}{\bf Q}_{11}=0$ and ${\rm Re}{\bf Q}_{22}=0$.
If we take the limit ${\bf q}=0$ and $\omega=2\Delta_0$, the equation ${\rm Re}{\bf Q}_{11}=0$ reduces to the gap equation \be -\frac{m}{4\pi a_s}=\int\frac{d^3{\bf k}}{(2\pi)^3}\big\{\frac{1}{2E}-\frac{1}{2\epsilon_{\bf k}}\big\}.\label{eq:gap}\ee Using the gap equation we can eliminate the term $m/4\pi a_s$ in ${\rm Re}{\bf Q}_{11}=0$ and finally obtain an equation as $\int\frac{d^3{\bf k}}{(2\pi)^3}\Big\{\frac{E+E'}{2EE'}\cdot\frac{EE'+\xi\xi'-\Delta_0^2}{\omega^2-(E+E')^2}+\frac{1}{2E}\Big\}$. In the limit of small ${\bf q}$, this equation can be rearranged as \bea (\omega^2-\frac{1}{3}v_F^2q^2-4\Delta_0^2)\int\frac{d^3{\bf k}}{(2\pi)^3}\frac{1}{2E(\omega^2-4E^2)},\label{eq:dispersion}\eea where $v_F$ is the Fermi velocity. This equation explicitly shows that there is a collective mode with dispersion relation as $\omega^2=\frac{1}{3}v_F^2q^2+4\Delta_0^2$. This is the Higgs mode. The gap is $2\Delta_0$. Analogously, we can obtain the dispersion relation of the gapless Goldstone mode as $\omega^2=\frac{1}{3}v_F^2q^2$ from equation ${\rm Re}{\bf Q}_{22}=0$.

However, as one approaches the unitary regime the particle-hole symmetry is gradually lost. The off-diagonal term ${\bf Q}_{12}$ is not negligible anymore. The amplitude fluctuation $\delta_a$ and $\delta_p$ will couple together. To investigate the behavior of the eigenmodes we evaluate the spectral weight function of $\delta_a$ and $\delta_p$, which can be calculated from the imaginary part of the propagator $<\delta_a^\ast\delta_a>$ and  $<\delta_p^\ast\delta_p>$ after the analytical continuation.
\bea &&A_{aa}(\omega, {\bf k})= \cr&&-\frac{1}{\pi}{\rm Im}\frac{{\bf Q}_{22}(\omega,{\bf k})}{{\bf Q}_{11}(\omega,{\bf k}){\bf Q}_{22}(\omega,{\bf k})-{\bf Q}_{12}(\omega,{\bf k}){\bf Q}_{21}(\omega,{\bf k})},\cr &&A_{pp}(\omega, {\bf k})= \cr&&-\frac{1}{\pi}{\rm Im}\frac{{\bf Q}_{11}(\omega,{\bf k})}{{\bf Q}_{11}(\omega,{\bf k}){\bf Q}_{22}(\omega,{\bf k})-{\bf Q}_{12}(\omega,{\bf k}){\bf Q}_{21}(\omega,{\bf k})}.\label{eq:A}\cr&&\eea
\begin{figure}[h]
\begin{center}
  \includegraphics[width=9cm]{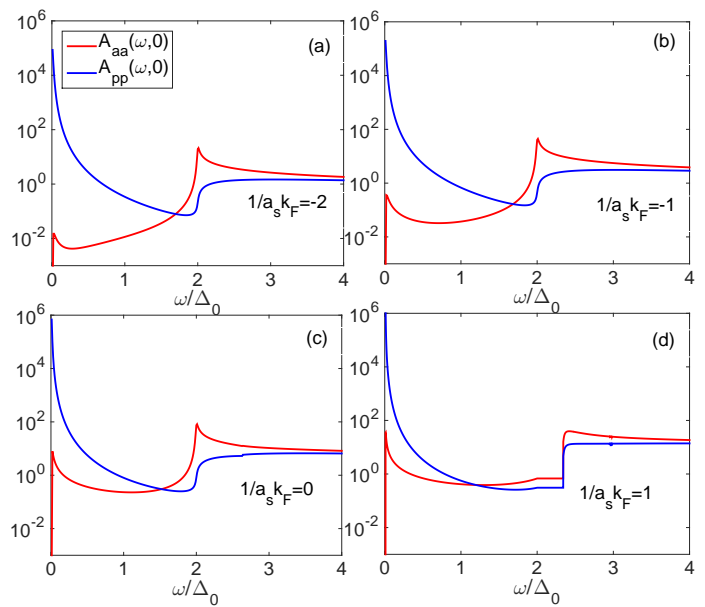}
  \caption{(Color online) The spectral weight functions $A_{aa}(\omega,0)$ and $A_{pp}(\omega,0)$ of amplitude and phase fluctuations at zero temperature. Graph (a), (b), (c) and (d) are for different scattering lengths $1/a_sk_F= $ -2, -1, 0, and 1.  }
  \label{fig:spectral0}
  \end{center}
 \end{figure}

In Fig. \ref{fig:spectral0} we demonstrate the spectral weight functions $A_{aa}(\omega, {\bf k})$ and $A_{pp}(\omega, {\bf k})$ for amplitude and phase fluctuations respectively at zero temperature. For simplicity we take the case of zero momentum. The four graphs are for different scattering lengths from the BCS to BEC side. We observe two features: (i) The spectral function of the phase fluctuation $A_{pp}(\omega, 0)$ only show one sharp peak at $\omega=0$, which is the gapless Goldstone mode. At higher frequency $\omega>2\Delta_0$ there is the two particle continuum. As we have discussed the phase and amplitude fluctuations strongly couple together in unitary regime and BEC side. Then the phase fluctuation field is a superposition of two eigen-modes: the Goldstone and Higgs mode. However, the Higgs mode doestn't manifest itself in $A_{pp}(\omega, 0)$. This means the probe couple to the phase fluctuation won't reveal any feature of Higgs mode in BEC-BCS crossover. (ii) The spectral function of the amplitude fluctuation $A_{aa}(\omega, 0)$ shows two peaks at the BCS limit and the unitary regime. One is the gapless Goldstone mode. The other one is the Higgs mode, which has a gap of $2\Delta_0$ as calculated in Eq. (\ref{eq:dispersion}). As one approaches the unitary regime from the BCS side the spectral weight transfer from Higgs to Goldstone mode is enhanced because the Lorentz invariance is lost as we discussed in Ref. \cite{Liu}. At the BEC limit the spectral weight of Higgs mode is further dimishied and finally overdamped by the two-particle continuum. However, around the unitary regime the spectral function $A_{pp}(\omega, 0)$ still possess a feature of gaped Higgs mode even though the Lorentz invariance is lost here.

\section{The spectral weight function at finite temperature\label{sec:T}}
 In Fig. \ref{fig:spectral0} we show that at zero temperature the spectral function $A_{aa}(\omega, 0)$ shows a clear peak of Higgs mode at the unitary regime. In this section we want to investigate if the sharp peak feature of the Higgs mode still exists at higher temperature around the unitary regime since in the present experiments the superfluid transition temperature can be reached in this region.

\begin{figure}[t]
 \includegraphics[width=0.48\textwidth]{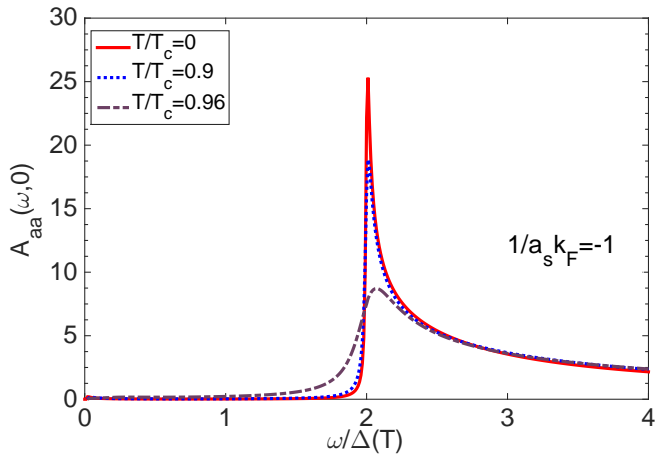}
  \caption{(Color online) The spectral weight function $A_{aa}(\omega, 0)$ at finite temperature. The scattering length is fixed at $1/a_sk_F=-1$. As the temperature increases the peak of Higgs mode gets broadened. The superfluid transition temperature $T_c$ at $1/a_sk_F=-1$ is $0.125T_F$}
\label{fig:spectralT} \end{figure}

 Assuming that the mean value of the Cooper pair field is $\Delta(T)$ at
finite temperature the effective action can be derived from Eq.
(\ref{eq:Z}) as $S_{\rm eff}=\beta
V\frac{\Delta(T)^2}{g}-\ln\det\hat G^{-1}$. Using the saddle point
condition $\partial S_{\rm eff}/\partial \Delta(T)=0$, we obtain the
gap equation as \bea -\frac{m}{4\pi a_s}=\int\frac{d^3{\bf
k}}{(2\pi)^3}\Big\{\frac{1-2f}{2E}-\frac{1}{2\epsilon_{\bf
k}}\Big\}.\label{eq:gapT}\eea To solve above equation for
$\Delta(T)$ we need to determine the chemical potential $\mu(T)$.
The total number of the particles is $N=-k_B T\partial S_{\rm
eff}/\partial \mu(T)$. Then it's straight forward to calculate the
number equation as \bea n=\int\frac{d^3{\bf
k}}{(2\pi)^3}\Big\{1-\frac{\xi}{E}(1-2f)\Big\}.\label{eq:numberT}\eea
Using Eq. (\ref{eq:gapT}) and  Eq. (\ref{eq:numberT}) the finite
temperature gap $\Delta(T)$ and chemical potential $\mu(T)$ can be
solved self-consistently. At finite temperature the spectral
functions $A_{aa}(\omega, {\bf k})$ and $A_{pp}(\omega, {\bf k})$
can still be calculated by Eq. (\ref{eq:A}) except that we use the
finite temperature matrix elements $Q_{ij}(\omega,{\bf k})$ and
$M_{ij}(\omega,{\bf k})$ in Eq. (\ref{eq:M}) and Eq. (\ref{eq:Q}).
The finite temperature spectral function of the amplitude
fluctuation $A_{aa}(\omega, {\bf k})$ at $1/a_sk_F=-1$ is
illustrated in Fig. \ref{fig:spectralT}. We investigate the case of
$1/a_sk_F=-1$ for two reasons. First, the superfluid transition
temperature is reachable here in the realistic experiments. Second,
our calculation of the critical temperature and $\Delta(T)$ is based
on the mean-field theory. It's not accurate close to the unitarity
and on the BEC side. However, at $1/a_sk_F=-1$ the values are close
to the Nozi\`{e}res Schmitt-Rink calculation \cite{NSR}. So it is an
appropriate point to study the finite temperature spectral
functions.
 We observe that as the
temperature is increased the peak of Higgs mode is broadened due to
the decay of the collective mode to the quasi-particles. At
temperature very close to $T_c$, for instance $T/T_c=0.96$, the
Higgs peak can get overdamped. However, for a wide range of
temperature from $T=0$ to $T/T_c=0.9$ the Higgs mode still presents
with a nice feature of sharp peak. This temperature range is
reachable in the realistic experiment of dilute Fermi gas. For
example, in MIT's experiment \cite{Zwierlein} the temperature can be
reduced to $T/T_F\simeq0.05$ after the evaporative cooling. The
superfluid transition temperature at $1/a_sk_F=-1$ is calculated as
$T_{c}=0.125T_{F}$ in our mean-field theory. This result
demonstrates an opportunity of the observation of Higgs mode in the
fermionic superfluid with tunable interaction.

\section{Conclusions\label{sec:con}}
In summary, we have investigated the observability of the Higgs mode
in BCE-BCS crossover. The superfluid order parameter is treated as a
spatially uniform mean-field plus phase and amplitude fluctuations.
We study the observability of the Higgs mode by calculating the
spectral weight function of the fluctuations. We emphasize that the
usual analysis of small frequency expansion \cite{Liu, Engelbrecht}
is not legitimate anymore in our mean-field study. In this work we
directly calculate the spectral functions with the complete integral
form of the matrix elements in Eq. (\ref{eq:M}). The zero
temperature study shows that the Higgs mode is totally overdamped by
the two particle continuum in the phase fluctuation spectral
function $A_{pp}(\omega,{\bf k})$ for the whole BEC-BCS crossover.
However, the signal of Higgs mode is clear in the amplitude
fluctuation spectral function $A_{pp}(\omega,{\bf k})$ even at the
unitary regime. At finite temperature, we show that in the
temperature region that present experiments can reach the Higgs mode
has a nice feature of sharp peak around the unitary region. This
suggests a potential observation of Higgs mode in the system of
dilute Fermi gases with tunable interactions.


\section{Acknowledgements} We thank Hui zhai and Shizhong Zhang for very useful
discussions. The work is supported by the Ministry of Science and Technology of China 973 program( No. 2015CB921300), National Science Foundation of China (Grant No. NSFC-1190020, 11334012), and the Strategic Pri- ority Research Program of CAS (Grant No. XDB07000000).


\begin{thebibliography}{Higgs}
\bibitem{Pekker2014} D. Pekker, and C. M. Varma, Annu. Rev. Condens. Matter Phys. \textbf{6}, 269 (2015).

\bibitem{Cern1} CMS collaboration, Phys. Lett. B \textbf{716}, 30 (2012).
\bibitem{Cern2} ATLAS collaboration, Phys. Lett. B \textbf{716}, 1 (2012).
\bibitem{Sooryakumar} R. Sooryakumar, and M. V. Klein, Phys. Rev. Lett. \textbf{45}, 660 (1980).
\bibitem{Littlewood1} P.B. Littlewood and C.M. Varma, Phys. Rev. B \textbf{26}, 4883 (1982).
\bibitem{Littlewood2} P.B. Littlewood and C.M. Varma, Phys. Rev. Lett. \textbf{47}, 811 (1981).

\bibitem{Ruegg} Ch. R\"{u}egg, B. Normand, M. Matsumoto, A. Furrer, D.F. McMorrow, K.W. Kramer, H.U. Gudel, S.N.
Gvasaliya, H. Mutka, and M. Boehm, Phys. Rev. Lett. \textbf{100}, 205701 (2008).

\bibitem{Matsunaga1} R. Matsunaga, Y. I. Hamada, K.
Makise, Y. Uzawa, H. Terai, Z. Wang, and R. Shimano, Phys. Rev.
Lett. 111, 057002 (2013).
\bibitem{Matsunaga2} R. Matsunaga, N. Tsuji, H. Fujita, A. Sugioka,
K. Makise, Y. Uzawa, H. Terai, Z. Wang, H. Aoki and R. Shimano,
Science \textbf{345}, 1145 (2014).

\bibitem{Endres} M. Endres, T. Fukuhara, D. Pekker, M. Cheneau, P.
Schaub, C. Gross, E. Demler, S. Kuhr and I. Bloch, Nature
\textbf{487}, 454-458 (2012).




\bibitem{Varma2002} C. M. Varma, J. Low Temp. Phys. \textbf{126} , 901 (2002).
\bibitem{Barlas2013} Y. Barlas, and C. M. Varma, Phys. Rev. B \textbf{87}, 054503 (2013).
\bibitem{Huber2008} S. D. Huber, B. Theiler, E. Altman, and G. Blatter, Phys. Rev. Lett. \textbf{100}, 050404 (2008).
\bibitem{Pollet2012} L. Pollet, and N. Prokof'ev, Phys. Rev. Lett. \textbf{109}, 010401 (2012).

\bibitem{Podolsky2011} D. Podolsky, A. Auerbach, and D. P. Arovas, Phys. Rev. B \textbf{84}, 174522 (2011).
\bibitem{Podolsky2012} D. Podolsky, and S. Sachdev, Phys. Rev. B \textbf{86}, 054508 (2012).
\bibitem{Gazit2013} S. Gazit, D. Podolsky, and A. Auerbach, Phys. Rev. Lett. \textbf{110}, 140401 (2013).
\bibitem{Rancom2014} A. Ran\c{c}on, and N. Dupuis, Phys. Rev. B \textbf{89}, 180501(R) (2014).
\bibitem{Liu} B. Liu, H. Zhai, and S. Zhang,  arXiv:1502.00431.
\bibitem{Cea} T. Cea, C. Castellani, G. Seibold, and L. Benfatto, Phys. Rev. Lett. \textbf{115}, 157002 (2015).


\bibitem{Engelbrecht} J. R. Engelbrecht, M. Randeria, and C. A. R. S\'{a} de Melo, Phys. Rev. B \textbf{55}, 15153 (1997).
\bibitem{Diener} R. B. Diener, R. Sensarma, and M. Randeria, Phys. Rev. A \textbf{77}, 023626 (2008).
\bibitem{NSR}P. Nozi\`{e}res, and S. Schmitt-Rink, J. Low Temp.
Phys. \textbf{59}, 195 (1985).
\bibitem{Zwierlein} M. W. Zwierlein, J. R. Abo-Shaeer, A. Schirotzek, C. H. Schunck and W. Ketterle, Nature,
\textbf{435}, 1047 (2005).

\end{thebibliography}
\end{document}